\documentclass[12pt]{article}
\usepackage{amsmath}
\usepackage{amssymb}
\usepackage{amsthm,amsxtra}
\usepackage{epsf}
\usepackage{eepic}
\usepackage{graphicx}
\newlength{\mytopmargin}
\newlength{\myleftmargin}
\usepackage[square,sort,comma,numbers]{natbib}
\newtheorem{lemma}{Lemma}
\newtheorem{prop}[lemma]{Proposition}

\numberwithin{equation}{section}

\begin{document}

\title{\Large\bf  Octonions in random matrix theory}
\author{Peter J. Forrester}
\date{}
\maketitle

\begin{center}
\it
Department of Mathematics and Statistics, ARC Centre of Excellence for Mathematical
 and Statistical Frontiers, The University of Melbourne, \\
Victoria 3010, Australia
\end{center}

\bigskip
\begin{center}
\bf Abstract 
\end{center}

The octonions are one of the four normed division algebras, together with the real, complex and quaternion number systems.
The latter three hold a primary place in random matrix theory, where in applications to quantum physics they are determined as the entries of ensembles of Hermitian random by symmetry considerations. Only for $N=2$ is there an existing analytic theory of Hermitian random
matrices with octonion entries. We use a Jordan algebra viewpoint to provide an analytic theory for $N=3$. We then proceed to consider the matrix structure $X^\dagger X$, when $X$ has random octonion entries. Analytic results are obtained from $N=2$, but are observed to break down in the $3 \times 3$ case.

\par
\bigskip
\noindent 
{\small 
}

\section{Introduction}

The classical matrix groups $O(N)$, $U(N)$, $Sp(2N)$ are unitary matrices with elements from the real, complex
and quaternion number fields respectively, the latter represented as $2 \times 2$ complex matrices. The first
two of these featured in the paper of A.~Hurwitz \cite{Hu97} introducing the notion of the invariant measure on
matrix groups. Hurwitz gave the specific form of the invariant measures in terms of Euler angle parameterisations and
computed the corresponding volumes. In so doing he arguably initiated the field of random matrix theory in mathematics
\cite{DF15}.

In theoretical physics ensembles of Hermitian random matrices with real entries were introduced by Wigner 
\cite{Wi57} as a model of the statistical properties of the highly excited spectra of heavy nuclei. Subsequently Dyson \cite{Dy62}
showed that Hermitian random matrices with real, complex and quaternion elements are in correspondence with quantum
systems possessing a time reversal symmetry $T$ with the property that $T^2=1$, no time reversal symmetry,
and a time reversal symmetry with the property $T^2 = -1$ respectively. The motivations for an ensemble theory can
be found, for example, in the introduction of the book by Porter \cite{Po65}, which includes many reprints of early papers in
the field, including those by Wigner and Dyson.

It was remarked in \cite{Dy62a} that real, complex and quaternion number systems are three of the four normed
division algebras.\footnote{Given Hurwitz's association with the beginnings of random matrix theory through \cite{Hu97}, it
is relevant to mention that it was Hurwitz  who proved this theorem \cite{Hu23}.} The fourth is the octonions.
In fact, taken literally, the octonions are incompatible with matrix algebra as they are not associative.
In a footnote \cite[Footnote 10]{Dy62a} Dyson makes this point, and goes on to say `We have tried and failed, to find
a natural way to fit octonions into the mathematical framework developed in this paper'.

In subsequent years there has been a number of papers relating to the eigenvalue problem for matrices with octonion entries;
see for example \cite{DM98,Ti00} and the book \cite{DM15}. Knowledge of some of these developments allowed the present author to specify an ensemble
of $2 \times 2$ Gaussian Hermitian random matrices with octonion entries \cite[\S 1.3.5]{Fo10}. These matrices can be represented as
$16 \times 16$ matrices with real entries, having 8 fold degenerate eigenvalues, and the resulting eigenvalue probability
density function (PDF) is proportional to
\begin{equation}\label{Ab}
e^{- c (\lambda_1^2 + \lambda_2^2)} | \lambda_2 - \lambda_1|^\beta
\end{equation}
with $\beta = 8$. In the case of $2 \times 2$ Gaussian Hermitian random matrices with real, complex and real
quaternion entries the functional form (\ref{Ab}) again specifies the eigenvalue PDF, now with $\beta = 1,2$ and 4 respectively.
It is furthermore the case that when represented as real matrices, the eigenvalues have degeneracies equal to these same values
of $\beta$.

On the other hand, it is known that representing $3 \times 3$ octonion matrices as $24 \times 24$ real
matrices generically gives six four fold degenerate eigenvalues
 \cite{Og81,DM98}, while for $N > 3$ there is no degeneracy at all \cite{Ni16}. One might then anticipate that there is some additional structure
 present for $N=3$, different from that when $N = 2$, and that all structure is lost beyond $N = 3$.
 Actually it is a classical result (see e.g.~\cite{DM15}) that $3 \times 3$ Hermitian matrices with octonion entries are distinguished as
 being Jordan algebras with the Jordan product rule 
 \begin{equation}\label{JP}
 A \circ B = {1 \over 2} (AB + BA).
\end{equation} 
  It is an objective of this paper to make
 use of the associated theory to specify a corresponding random matrix ensemble with eigenvalue PDF proportional to 
 \begin{equation}\label{TX}
 \prod_{l=1}^3 e^{- c \lambda_l^2} \prod_{1 \le j < k \le 3} | \lambda_k - \lambda_j |^8.
 \end{equation}
 
 Before random matrix ensembles of real symmetric matrices were isolated by Wigner for their interest in theoretical physics,
 positive definite matrices
 \begin{equation}\label{WX}
 W = X^\dagger X,
 \end{equation}
 with $X$ and $n \times N$ $(n \ge N)$ real standard Gaussian random matrix had been extensively studied in the mathematical
 statistics literature for their relevance to multivariate statistical analysis (see e.g.~the texts \cite{Mu82,An04}).  
 The pioneering work in that field was due to Wishart \cite{Wi28}, giving rise to the name Wishart matrices and Wishart distributions
 for random matrices formed with the structure (\ref{WX}).
 Subsequently these studies
 were extended to include the cases that the Gaussian matrix $X$ has complex, or quaternion, entries; see
 e.g.~\cite[Ch.~3]{Fo10}. The question then arises as to properties of the random matrix structure (\ref{WX}) in the case that
 $X$ contains octonion entries. We will exhibit two constructions for $N = 2$ that lead to an eigenvalue PDF
 proportional to
 \begin{equation}\label{3X} 
 (\lambda_1 \lambda_2)^a e^{-c (\lambda_1 + \lambda_2)} ( \lambda_2 - \lambda_1)^8, \qquad \lambda_1, \lambda_2 > 0,
 \end{equation}
 for particular $a$ (as in (\ref{TX}) $c$ is simply a scale factor). However, these both breakdown for $N = 3$, leaving us without
 a construction involving octonions of the $N = 3$ generalisation of (\ref{3X}).
 
 We begin in Section \ref{S2} with a review of the $2 \times 2$ Hermitian octonionic eigenvalue problem, and how it leads to
 the PDF (\ref{Ab}) with $\beta = 8$. We then proceed to exhibit how notions from the theory of Jordan algebras can be used to
 realise (\ref{TX}) in an octonionic setting. In Section 3 we introduce the matrix structure (\ref{WX}), specialising first to the case
 $N=2$. Rectangular matrices $X$ of size $n \times 2$ with octonion entries are shown to lead to (\ref{3X}), as is the case that
 $X$ is a $2 \times 2$ triangular random matrix with positive real entries on the diagonal and a single off diagonal octonion
 entry. We observe that the analogous construction in the case $N=3$ does not lead to positive definite matrices, leaving us
 without a constructive theory of $3 \times 3$ Wishart matrices with octonion entries.
 
\section{Hermitian random matrices with octonion entries}\label{S2}
\subsection{Preliminaries}
One recalls (see e.g.~\cite{DM15}) that quaternions are a non commutative four dimensional algebra with units $\{1,i,j,k\}$,
having the properties that $i^2=j^2=k^2 = -1$, $\{i,j,k\}$ anti-commute in pairs, and $ijk=-1$. The units can be represented as
the $2 \times 2$ complex matrices
\begin{equation}\label{2.1}
\begin{bmatrix}1 & 0 \\
0 & 1 \end{bmatrix}, \quad \begin{bmatrix}i & 0 \\
0 & -i \end{bmatrix},  \quad \begin{bmatrix}0 & 1 \\
-1 & 0 \end{bmatrix},   \quad \begin{bmatrix}0 & i \\i & 0 \end{bmatrix},
 \end{equation}
 respectively. In random matrix theory, reference to matrices with quaternions entries actually refers to complex matrices
 with entries consisting of $2 \times 2$ blocks of the form (\ref{2.1}). A fundamental property of such complex matrices is that 
 they have the property $X = Z_{2N} \bar{X} Z_{2N}^{-1}$, where $Z_{2N} = 1_N \otimes \begin{bmatrix} 0 & -1 \\ 1 & 0 \end{bmatrix}$.
 Consequently, if ${\psi}$ is an eigenvector with eigenvalue $\lambda$, so is $Z_{2N}  {\bar{\psi}}$, and these two eigenvectors are
 furthermore orthogonal. Now write each entry in (\ref{2.1}) in the form $a+ib$, and map the complex $2 \times 2$ matrices
 to $4 \times 4$ matrices with real entries according to the 
standard representation of the complex number $a + i b$ as the 
 $2 \times 2$ real matrix $\begin{bmatrix} a & b \\ -b & a \end{bmatrix}$. Applying the same mapping to $\psi$ gives two columns of
 orthogonal real eigenvectors, as does $Z_{2N}  {\bar{\psi}}$, thus now giving a four fold degenerate spectrum.
 
 Let $p_1, p_2$ be quaternions. The octonion algebra consists of elements of the form $p_1 + p_2 l$, with $p_2 l$ 
 algebraically independent of $p_1$, so that the (real) octonions are an eight dimensional algebra over the reals with units
 $$
1, \: \: e_1 := i, \: \: e_2 := j, \: \: e_3 := k, \: \:
e_4 := l, \: \: e_5 := il, \: \: e_6 := jl, \: \: e_7 := kl.
$$
Addition and multiplication are defined by
\begin{equation}\label{1.mo1}
a+b = (p_1 + q_1) + (p_2 + q_2) l, \qquad
a b = (p_1 q_1 - \bar{q}_2 p_2) + (q_2 p_1 + p_2 \bar{q}_1) l,
\end{equation}
respectively. 
In general
$$
a(bc) \ne (ab)c
$$
(for example with $a=e_5$, $b=e_6$, $c=e_7$ we have
$e_5(e_6 e_7) = - (e_5 e_6) e_7$), so unlike the quaternions the
octonions are not associative.

But other properties of the quaternions are maintained.
Let $a=  a_0 + \sum_{j=1}^7 a_j e_j$, and define $\bar{a}=  a_0 - \sum_{j=1}^7 a_j e_j$. Then
$\overline{ab} = \bar{b} \bar{a}$ and thus with
\begin{equation}\label{1.toc1}
|a| := \sqrt{a \bar{a}} = \sqrt{\bar{a} a} =  \sqrt{a_0^2 + a_1^2 + \cdots + a_7^2},
\end{equation}
we have 
\begin{equation}\label{1.toc2}
|a b| = |a| \, |b|.
\end{equation}
It is also true that each $a \ne 0$ has a unique inverse specified by
\begin{equation}\label{1.toc3}
a^{-1} = \bar{a}/(\bar{a} a).
\end{equation}
The properties (\ref{1.toc1})--(\ref{1.toc3}) say that the real octonions 
are a normed division algebra. 

We know that the matrices (\ref{2.1}) are a complex matrix representation of the algebra of quaternions, and we discussed
in the paragraph below how to convert it to a real matrix representation. Due to the octonions being non-associative, there
can be no analogue of these representations. On the other hand, if we consider instead right and left multiplication separately
there do exist well defined (real) matrix representations (see e.g.~\cite{Ti00}, \cite[Prop.~1.3.6]{Fo10}). The main result for
present purposes is that with $x=  x_0 + \sum_{j=1}^7 x_j e_j$ a quaternion, and $\vec{x} = (x_0, \dots, x_7)^T$ 
the column vector formed from the coefficients, one has $\vec{a x} = \omega (a) \vec{x}$, where
\begin{equation}\label{pa}
\omega(a) = 
 \left [ \begin{array}{cccccccc}
a_0 & - a_1 & - a_2 & - a_3 & - a_4 & -a_5 & -a_6 & -a_7 \\
a_1 & a_0 & -a_3 & a_2 & -a_5 & a_4 & a_7 & -a_6 \\
a_2 & a_3 & a_0 & -a_1 & -a_6 & -a_7 & a_4 & a_5 \\
a_3 & -a_2 & a_1 & a_0 & -a_7 & a_6 & -a_5 & a_4 \\
a_4 & a_5 & a_6 & a_7 & a_0 & -a_1 & -a_2 & -a_3 \\
a_5 & -a_4 & a_7 & -a_6 & a_1 & a_0 & a_3 & -a_2 \\
a_6 & -a_7 & -a_4 & a_5 & a_2 & -a_3 & a_0 & a_1 \\
a_7 & a_6 & -a_5 & -a_4 & a_3 & a_2 & -a_1 & a_0
\end{array} \right ] .
\end{equation}

\subsection{The case $N=2$}
A $2 \times 2$ Hermitian matrix with octonian elements must have the form
\begin{equation}\label{AA}
H = \begin{bmatrix}a & b \\
\bar{b} & c \end{bmatrix}
\end{equation}
with $a$ and $c$ with the property that $a = \bar{a}$ and $b = \bar{b}$, and thus having only the coefficient of
the unit 1 as nonzero (the usual terminology is that $a$ and $b$ are then real)
For the quaternion case, we know that to build a theory in which (\ref{AA}) has real eigenvalues, it is necessary
to make use of the complex representation (\ref{2.1}). But we have just revised that octonions necessarily have no analogue
of such a representation, and instead a matrix representation is restricted to right (or left) multiplication, which in the former
case is described explicitly by (\ref{pa}). This suggests to study $16 \times 16$ real symmetric matrices of the form
\begin{equation}\label{1.octa}
 \omega(H ) =
\left [ \begin{array}{cc} a{\mathbb I}_8 &  \omega(b) \\
 (\omega(b))^T & c{\mathbb I}_8 \end{array} \right ].
\end{equation}

As noted in \cite{Ti00}, the characteristic polynomial of (\ref{1.octa}) factorises according to
\begin{equation}\label{1.octaA}
\det ( \omega(H) - \lambda \mathbb I_{16}) = ((a - \lambda) (c - \lambda) - b \bar{b})^8,
\end{equation}
showing that each eigenvalue is eightfold degenerate. Following \cite[\S 1.3.5]{Fo10}, to now construct a theory
of random Hermitian matrices, one specifies that the eigenvalue PDF of $H$ is proportional to $e^{- (\tilde{c}/8) {\rm Tr} \, H^2}$.
This is equivalent to requiring that $a$ and $c$ in (\ref{1.octa}) have distribution given by the Gaussian 
N$[0,1/\sqrt{2\tilde{c}}]$, and each real coefficient in $b$ (there are eight in total) is independently distributed according to the
Gaussian N$[0,1/2\sqrt{\tilde{c}}]$. In this circumstance, $b \bar{b}$ consists of the sum of eight independent Gaussians of this latter
specification, and as such has distribution given by $\Gamma[4,1/2\tilde{c}]$, where  $\Gamma[\kappa, s]$ refers to the Gamma
distribution with shape $\kappa$ and scale $s$.

For definiteness, let us choose $\tilde{c}=1/2$, in which case N$[0,1/\sqrt{2\tilde{c}}]$ corresponds to a standard Gaussian.
In general, for a $2 \times 2$ real random matrix of the form (\ref{AA}) with $a, c$ distributed as independent standard
Gaussians, and $b$ distributed as the Gamma distribution $\Gamma[\kappa, 1]$, the eigenvalue PDF can readily be
computed to be proportional to (\ref{Ab}) with $c=1/2$ and $\beta = 2\kappa$ \cite{DE02}, \cite[Prop.~1.9.4 with $N=2$]{Fo10}.
Hence we have the following result for the Gaussian ensemble based on the structure (\ref{1.octa}) \cite[\S 1.3.5]{Fo10}.

\begin{prop}\label{P1}
Consider real symmetric matrices of the form (\ref{1.octa}). Let $a, b$ and the eight independent components of 
$c$ be distributed as standard Gaussians, and Gaussians {\rm N[0,$1/\sqrt{2}]$} respectively. The two distinct eigenvalues have
PDF proportional to (\ref{Ab}) with $c=1/2$ and $\beta = 8$.
\end{prop}

We remark that Li \cite{Li15} has recently generalised the above model of Gaussian $2 \times 2$ Hermitian matrices with
octonion entries to a model where the entries follow Brownian motion, extending the work \cite{BZ16}, which provides
a different derivation of Proposition \ref{P1}.

\subsection{The case $N=3$}

As already commented, forming a $3 \times 3$ Hermitian matrix with octonion elements, then converting it to a
$24 \times 24$ real symmetric matrix according to the prescription (\ref{1.octa}), gives matrices with generically
six independent eigenvalues, each of multiplicity four  \cite{Og81,DM98}. Thus it is not possible to realise the
eigenvalue PDF (\ref{TX}) in this way. Instead, one appeals to the Jordan algebra structure of the set of 
$3 \times 3$ Hermitian random matrices (see e.g.~\cite{Ba02j,DM15}). As made clear in \cite{DM99}, the correct way
to think of the eigenvalue problem is as the eigen-matrix equation
\begin{equation}\label{EP}
H \circ \Omega = \lambda \Omega,
\end{equation}
with the operation $\circ$ specified by (\ref{JP}). The $3 \times 3$ matrices with octonion entries $\Omega$ can be
chosen to have the projector property $\Omega \circ \Omega = \Omega^2 = \Omega$. A fundamental consequence of the Jordan algebra
structure of the set of $3 \times 3$ Hermitian matrix with octonion elements is that the equation (\ref{EP}) permits
three real eigenvalues, $\lambda_1, \lambda_2, \lambda_3$ say. For each eigenvalue there is a corresponding eigen-matrix $P_i$,
each of which is a projector (idempotent) and mutually orthogonal in the sense that 
$P_i \circ P_j = 0$, with 0 the $3 \times 3$ zero matrix, for distinct pairs. In terms of its eigenvalues and eigen-matrices the Hermitian
matrix
$H$ has the decomposition
\begin{equation}\label{EPa}
H  = \lambda_1 P_1 + \lambda_2 P_2 + \lambda_3 P_3.
\end{equation}

Most significant for present purposes is that eigenvalues $\lambda$ in (\ref{EP}) satisfy the characteristic equation
(see e.g.~\cite{Ma94,DM99})
\begin{equation}\label{EP1}
\lambda^3 - ({\rm Tr} \, H) \lambda^2 + \sigma(H) \,  \lambda - (\det H)  = 0,
\end{equation}
where,  writing $H = [X_{jk}]_{j,k=1,\dots,3}$ with $X_{jj} = x_{jj} I_8$, and $X_{ij} = x_{ij}$, $X_{ji} = \bar{x}_{ij}$ 
 for $i < j$,
 \begin{align}\label{2.13}
 {\rm Tr} \, H  & = x_{11} + x_{22} + x_{33} \nonumber \\
 \sigma (H) & = x_{11} x_{22} + x_{11} x_{33} + x_{22} x_{33} - |x_{12}|^2 - |x_{13}|^2 - |x_{23}|^2  \nonumber \\
 \det H & = x_{11} x_{22} x_{33} + 2 \, {\rm Re} \, \bar{x}_{13} ( x_{12} x_{23})  -
 x_{33} |x_{12}|^2 - x_{22} | x_{13}|^2 - x_{11} |x_{23}|^2,
 \end{align}
and in the final expression ${\rm Re} \, x = (x + \bar{x})/2$.  Thus one has the familiar formulas
$$
 {\rm Tr} \, H = \lambda_1 +  \lambda_2 + \lambda_3, \qquad
 \det H =   \lambda_1 \lambda_2 \lambda_3.
 $$
 We note too for future purposes that the decomposition (\ref{EPa}) together with the properties of the $P_i$ give the
 further familiar formula
\begin{equation}\label{EP2} 
 {\rm Tr} \, H^2 = \sum_{i=1}^3 \lambda_i^2 .
\end{equation} 

Let $F_4$ be the set of traceless $3 \times 3$ Hermitian matrices with octonian entries, generated by  matrices from this
same set, and further constrained to have
only one nonzero off diagonal entry, and to square to the identity.
As is well known (see \cite[Appendix]{DM99})
any $3 \times 3$ Hermitian matrix with octonion elements $H'$ can be diagonalised in terms of these matrices by an expansion
$$
H' = \xi \, {\rm diag} \, (\lambda_1, \lambda_2, \lambda_3) \, \xi^\dagger
$$
for eigenvalues $\lambda_1,\lambda_2,\lambda_3$ and some $\xi \in F_4$, where $\xi = \xi_3 \xi_2 \xi_1$ with each $\xi_i$
belonging to a different generator, and multiplication carried out according to the order $\xi_3 (\xi_2 (\xi_1 H'  \xi_1) \xi_2) \xi_3$.
From a theorem of Farout and
Koranyi \cite[Th.~VI.2.3 with $d=8$]{FK94} we know that the  measure $(dH')$ of the independent parts of the Hermitian matrix $H'$ decomposes in terms of 
the invariant measure for $F_4$, $(dF_4)$, and the eigenvalues according to 
\begin{equation}\label{EP2} 
 (dH') = c \prod_{1 \le j < k \le 3} (\lambda_k - \lambda_j)^8 \, d\lambda_1 d\lambda_2 d\lambda_3 (dF_4),
 \end{equation}
where $c$ is a constant.
Combining this with (\ref{EP2}), we obtain the following analogue of Proposition \ref{P1}.

 \begin{figure}[t]
 \includegraphics[scale=0.8]{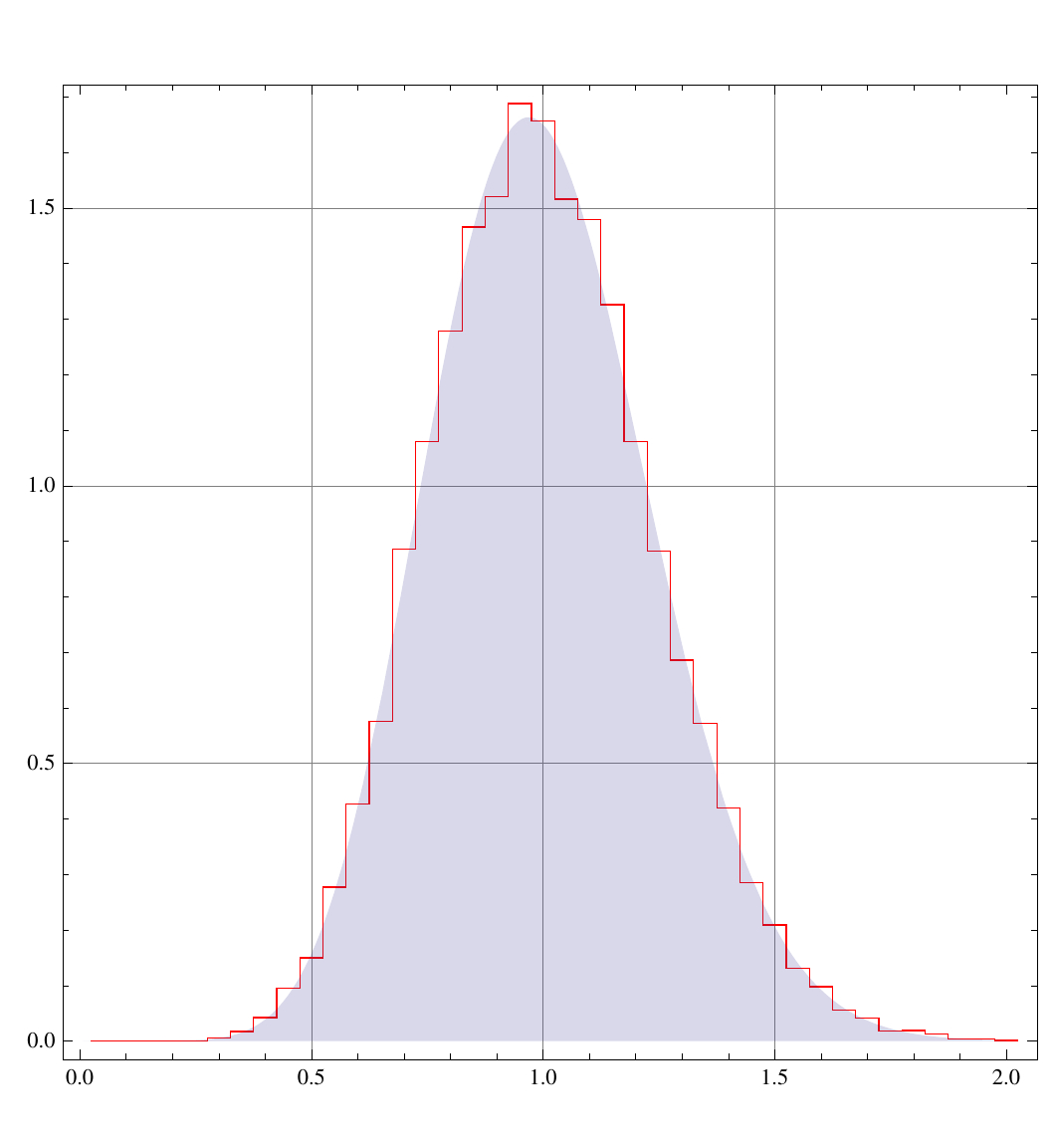} 
 \caption{\label{F1}Histogram of the spacing  distribution variables $\lambda_2 - \lambda_1$ and
 $\lambda_3 - \lambda_2$ computed from the roots of (\ref{EP1}) with the random variables of Proposition \ref{P2},
 and scaled to have mean unity, compared against the Wigner surmise (\ref{EP3}) with $\beta = 8$.
\label{F1}}
\end{figure}

\begin{prop}\label{P2}
Let $\alpha_i$, $i=1,2,3$ be independent standard Gaussians. Let $x_{ij}^{(s)}$ be, for $1 \le i < j \le 3$ and $s=0,1,\dots,7$,
independent real Gaussians {\rm N[0,1/$\sqrt{2}$]}. From the $x_{ij}^{(s)}$, form real octonions
$x_{ij} = x_{ij}^{(0)} + \sum_{s=1}^7  x_{ij}^{(s)} e_s$, and define the $3 \times 3$ Hermitian matrix with octonion elements
$H  = [X_{jk}]_{j,k=1,\dots,3}$ with $X_{jj} = \alpha_{j} I_8$, and $X_{ij} = x_{ij}$, $X_{ji} = \bar{x}_{ij}$ 
 for $i < j$.
The three eigenvalues of $H$ as determined by the solution of the characteristic equation (\ref{EP1}) have
PDF proportional to (\ref{TX}) with $c=1/2$ and $\beta = 8$.
\end{prop}

It is of interest to compare the prediction of Proposition \ref{P2} against numerical simulation; see Figure \ref{F1}. For this
we make  use of an analytic approximation for the unfolded (normalised to have mean unity) spacing between eigenvalues
in the Gaussian $\beta$ ensemble for general $N$ 
(i.e.~the PDF (\ref{Ab}) generalised to $N$ eigenvalues analogous to (\ref{TX}) for $\beta = 8$ and $N=3$) known 
as the Wigner surmise (see e.g.~\cite[Exercises 8.1 q.3]{Fo10}),
\begin{equation}\label{EP3} 
p_\beta^{\rm W} = {1 \over C} s^\beta e^{- \tilde{c} s^2}, \qquad
C_\beta = \Big ( {1 \over \beta \tilde{c}} \Big )^{(\beta+1)/2} 2^\beta \Gamma((\beta+1)/2), \: \: 
\tilde{c} = \bigg ( {\Gamma(\beta/2+1) \over \Gamma(\beta/2+1/2) } \bigg )^2.
 \end{equation}
 This is the exact distribution of the spacing variable $s = \lambda_2 - \lambda_1$ in (\ref{Ab}), scaled so that the 
 mean spacing is unity. We will use it as an approximation to the distribution of the spacing variables $s = \lambda_3 -
 \lambda_1$, $s = \lambda_2 - \lambda_1$ in  (\ref{TX}), which is not known exactly. The justification is that it is well known that the
 Wigner surmise is an accurate approximation, even in the case of the large $N$ limit. For example, with
 $\beta = 4$, from knowledge of the variance, skewness and kurtosis at $\beta = 1$ of the next neighbour spacing distribution from
 \cite[Table 1]{Bo09} or \cite[Table 8.14]{Fo10}, the fact that the nearest neighbour spacing at $\beta = 4$ 
 has the same variance divided by 4, and the same skewness and kurtosis, tells us that to 5 decimal places these statistical
 quantities have the values $0.104091$, $0.34939$ and $0.2858$ respectively. The Wigner surmise (\ref{EP3}) gives for these quantities
 the values 0.10447, 0.35939, 0.03698. 
 
 \section{Wishart matrices with octonion entries}
 \subsection{The case $N=2$}
 Let $X$ in (\ref{WX}) be an $n \times 2$ matrix with each element an octonion, and write $X = [ \vec{x}_1 \: \vec{x}_2 ]$ where
 $\vec{x}_i$ ($i=1,2$) is an $n \times 1$ column vector. Then $W$ is the $2 \times 2$ Hermitian matrix
 \begin{equation}\label{WP}
 W = \begin{bmatrix} \vec{x}_1^\dagger \vec{x}_1 &  \vec{x}_1^\dagger \vec{x}_2 \\
  \vec{x}_2^\dagger \vec{x}_1 &  \vec{x}_2^\dagger \vec{x}_2 \end{bmatrix}.
 \end{equation}
 Since the Cauchy--Schwarz inequality 
 \begin{equation}\label{CS}
  | \vec{x}_1^\dagger \vec{x}_2 |^2 \le    | \vec{x}_1^\dagger \vec{x}_1 | \,  | \vec{x}_2^\dagger \vec{x}_2 |
  \end{equation}
 remains valid for vectors with octonion entries (see e.g.~\cite[Proof of Lemma 1]{Kr98}),
  using the notation on the RHS of (\ref{AA}) for the entries of $W$, it follows that
  \begin{equation}\label{CS} 
a, c \ge 0, \qquad a c - |b|^2 \ge 0.
 \end{equation}
 With the eigenvalues of the $16 \times 16$ real symmetric matrix $\omega(W)$ being given by the roots
 of the quadratic on
 the RHS of (\ref{1.octa}) each with multiplicity 8, the inequalities (\ref{CS}) tell us immediately that the two roots of the quadratic
 are non-negative.
 
 To determine the PDF of the roots, with $W = X^\dagger X$ and the PDF for $X$  proportional to $e^{-(1/2) {\rm Tr} \, X^\dagger X}$,
 we want to determine the function $J(W)$ such that the PDF of $W$ is proportional to $J(W) e^{- (1/2) {\rm Tr} \, W}$.
 For this we use a method based on functional equations due to Rasch \cite{Ra48} in the real case,
 and popularised by Olkin \cite{Ol00,Ol02}. Its generalisation to the complex and quaternion cases was previously
 given in \cite[Exercises 3.2 q.6]{Fo10}.
 
 \begin{prop}\label{P3}
 Let $W = X^\dagger X$, where $X$ is an $n \times 2$ matrix with octonion entries, and let $X$ have PDF proportional to
 $F(X^\dagger X)$. With the entries of $W$ written according to the RHS of (\ref{AA}),
 define $\det W = ac - |b|^2$. We have that the PDF of $W$ is proportional to  $J(W) F(W)$ with
  \begin{equation}\label{J}
  J(W) = (\det W)^{4n - 5}.
  \end{equation}
  \end{prop}
  
  \noindent
  Proof. \quad
  With $V$ a $2 \times 2$ Hermitian matrix with octonion elements, and $B$ a general $2 \times 2$ matrix with octonian entries,
  let $W = B^\dagger V B$. Consideration of $B$ consisting of a product of elementary matrices (see e.g.~\cite[Exercises 1.3 q.2]{Fo10})
  shows
  that the PDF of $V$ is
 \begin{equation}\label{U1}
( \det B^\dagger B)^5 J( B^\dagger V B) F( B^\dagger V B),
 \end{equation}
or equivalently $(d W) = ( \det B^\dagger B)^5 (d V)$ (the notation $(dW)$, and similarly $(dV)$, denotes the products of the
independent differentials of the two diagonal elements, and the 8 independent real coefficients of the independent off diagonal entry).

Next let $X = Y B$, so that $Y$ is of size $n \times 2$, and furthermore requite that $V = Y^\dagger Y$. Then, adapting the method
of derivation of \cite[Prop.~3.2.4]{Fo10} shows
$(d X) = (\det B^\dagger \det B)^{4n}$, telling us that the PDF of $Y$ is 
$F(B^\dagger Y^\dagger Y B)  (\det B^\dagger \det B)^{4n}$ and hence that of $V$ us
 \begin{equation}\label{U2}
 (\det B^\dagger  B)^{4n}  J(V) F( B^\dagger V B).
 \end{equation}
 
 Equating  (\ref{U1}) and (\ref{U2}) then setting $V = \mathbb I_2$ shows that we must have
 $J( B^\dagger  B) = J(\mathbb I_2) (\det B^\dagger  B)^{4n-5}$. Since $J(\mathbb I_2)$ is a constant,
 this implies the result. \hfill $\square$
 
 \medskip
 
 Analogous to (\ref{EP2}), for $2 \times 2$ Hermitian matrix with octonian entries $\{W\}$, and
 denoting the set of $2 \times 2$ Hermitian orthogonal matrices with octonian entries diagonalising
 $W$ by $\{\tilde{F}\}$, from \cite[Eq.~(1.34)]{Fo10} we have
 \begin{equation}\label{EP2a}
 (d W) = (\lambda_2 - \lambda_1)^8 d \lambda_1 d \lambda_2 (d \tilde{F}).
 \end{equation} 
 With $W = X^\dagger X$ and the PDF for $X$  proportional to $e^{-(1/2) {\rm Tr} \, X^\dagger X}$, it follows
 from Proposition \ref{P3} that the PDF of $W$ is proportional to $ (\det W)^{4n - 5}  e^{- (1/2) {\rm Tr} \, W}$. Now changing
 variables to the eigenvalues (which from the discussion below (\ref{CS}) are non-negative) and the diagonalising matrices $\{ 
 \tilde{F} \}$ using (\ref{EP2a}) provides us with the functional form of the eigenvalue PDF.
 
 \begin{prop}\label{P4} 
  Let $W = X^\dagger X$, where $X$ is an $n \times 2$ matrix with random octonian entries, the eight independent components in
  each being distributed as standard Gaussians. The $16 \times 16$ real symmetric matrix $\omega(W)$ has two eight fold degenerate
  eigenvalues, and their PDF is proportional to (\ref{3X}) with $a = 4n - 5$ and $c=1/2$.
  \end{prop}
  
  The distribution of the smallest eigenvalue implied by Proposition \ref{P4} is
  \begin{align}\label{3.8}
  p(\lambda) & = {2 \over C} e^{- c s} s^{4n-5} \int_s^\infty e^{- c \lambda} (s - \lambda)^8 \lambda^{4n - 5} \, d \lambda \nonumber \\
  & =  {2 \over C} e^{- 2 c s} s^{4n - 5} \int_0^\infty e^{- c x} x^8 (s + x)^{4n - 5} \, dx.
  \end{align}
  Here $C$ is the the normalisation, chosen so that $\int_0^\infty p(\lambda) \, d \lambda = 1$, and $c=1/2$.
   For any integer $n \ge 2$,
  the integral in the final line of (\ref{3.8}) is a polynomial in $s$ of degree $4n - 5$ with coefficient of $s^l$ equal to
  $$
  c^{- (4 n + 3)+l} \binom{4n-5}{l} (3 + 4n - l)!.
  $$
  
\begin{figure}[t]
 \includegraphics [scale=0.85]{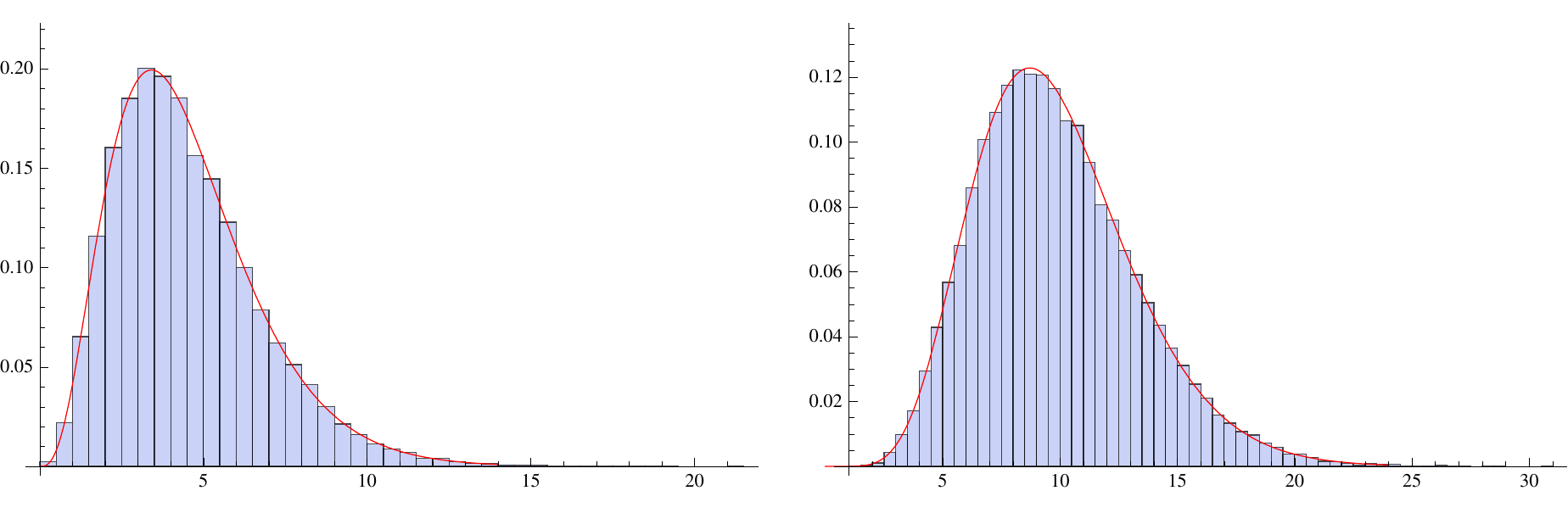}
 \caption{\label{F2}Histogram of the smallest eigenvalue PDF for $2 \times 2$ random matrices $X^\dagger X$, with each entry of the $n \times 2$ matrix $X$ a standard Gaussian octonion ($n=2$ and $n=3$ respectively) generated from a simulation, plotted against the
 theoretical predictions.}
\end{figure}

  In Figure \ref{F2} we compare this theoretical prediction in the cases $n=2$ and $n=3$ against histograms obtained from the
  numerical determination of the eigenvalues of the matrices $X^\dagger X$, with each entry of the $n \times 2$ matrix $X$ a standard Gaussian octonion.
  For the numerical computation, we can either simply directly solve the quadratic on the RHS of (\ref{1.octaA}), or seek instead
  the eigenvalues of the  eight fold degenerate $16 \times 16$ real symmetric matrix $\omega(W)$.
    
There is an alternative construction of octonion random matrices leading to the PDF (\ref{3X}). Instead of the structure $X^\dagger X$,
with $X$ a $n \times 2$ random matrix, consider instead $T^\dagger T$ with $T$ a $2 \times 2$ upper triangular matrix,
having its two diagonal entries real, and its off diagonal entry an octonion. In the real and complex cases, it is well known that
the existence of such a decomposition --- often called the Cholesky decomposition --- is equivalent to the matrix being positive
definite. In the present setting, with $W = T^\dagger T$, a straightforward computation generalising \cite[Prop.~3.2.6]{Fo10}
(see also \cite[Eq.~(2)]{Di16}) shows
  \begin{equation}\label{3.9}
  (d W) = 2^2 t_{11} t_{22}^9 (d T).
\end{equation}
This in turn tells us that
  \begin{equation}\label{3.10}
  (\det W)^a e^{- {\rm Tr} \, W/2} (d W) \propto t_{11}^{2a+1} t_{22}^{2a+9} e^{- (t_{11}^2 + t_{22}^2 + |t_{12}|^2)/2}.
 \end{equation}
 We can read off from this the specification on the nonzero elements of $T$ such that  the eigenvalue PDF of
 $W$ is given by (\ref{3X}).
 
 \begin{prop}\label{P5} 
  Let $W = T^\dagger T$, with $T = \begin{bmatrix} t_{11} & t_{12} \\  0 & t_{22} \end{bmatrix}$.
  Choose the off diagonal element $t_{12}$ to be an octonion with its eight independent components distributed as standard Gaussians.
  Choose the diagonal elements to be positive and real, and specified by choosing $t_{11} = s_1^{1/2}$, $t_{22} = s_2^{1/2}$
  with $s_1, s_2$ having the Gamma distributions $\Gamma[a+1,2]$ and $\Gamma[a+5,2]$ respectively. The matrix $W$ then has
  eigenvalue PDF proportional to (\ref{3X}) with $c=1/2$.
  \end{prop}
  
  Note that unlike the construction in Proposition \ref{P4}, which implies the specific form for the parameter $a$ in  (\ref{3X}),
  $a= 4n - 5$, with $n \ge 2$ and an integer,
 the construction of Proposition \ref{P5} is well defined for all $a > -1$.

  \subsection{The case $N=3$}
  
  For $N=2$ we have seen that the determinant of $2 \times 2$ Hermitian matrices with octonian entries is defined
  according to the usual formula.  When it comes to $N=3$, we know from (\ref{2.13}) that the order of multiplying 
  the three independent octonians plays a role in the definition. With  $W = T^\dagger T$, this in turn destroys familiar properties like
  $\det W = | \det T|^2$.
  
  A dramatic illustration of this last point can be had by using simulation to compute the sign of $\det  T^\dagger T$, with
    $$
  T = \begin{bmatrix} t_{11} & t_{12} & t_{13} \\
  0 & t_{22} & t_{23} \\
  0 & 0 & t_{33} \end{bmatrix},
  $$
  where each diagonal entry is positive and real, given by $t_{ii} = s_i^{1/2}$,
  $s_i$ having the Gamma distributions $\Gamma[a + 4(i-1),2]$, and each $t_{ij}$, $i \ne j$ an octonion
  with independent components distributed as standard Gaussians. This is the natural $N=3$ generalisation of
  the random matrix $W$ in Proposition \ref{P5}. Of course $\det T = \det T^\dagger = \prod_{i=1}^3 t_{ii}$ according to 
  the definition (\ref{2.13}). But generating 10,000 $3 \times 3$ random matrices $W$ gave over 5,500 having a negative
  determinant. As a consequence, this prescription can no longer be used to generate positive definite matrices in the sense
  that all the eigenvalues are positive.
  
  Empirically, we have observed that the Jordan product
  $$
  {1 \over 2} ( T^\dagger T + T T^\dagger),
  $$
with $T$ defined as in the previous paragraph  has all but a very small fraction of its eigenvalues positive; typically only
3 out 10,000 trials giving a negative eigenvalue.  But with this fraction being nonzero, still we remain without a construction of random $3 \times 3$ positive
definite matrices with octonion entries, and in particular without a way to generate eigenvalues with PDF given by
 the $N=3$ analogue of
(\ref{3X}) using such matrices, or equivalently without a prescription of matrices distributed according to the LHS of (\ref{3.10})
for $N=3$.
The latter  is the Wishart distribution with  covariance
matrix proportional to the identity.  This is somewhat ironic as a number of studies highlight the natural place for this distribution in the context of exponential
models and symmetric cones \cite{Je88,Ma94h,AW04}.

\subsection*{Acknowledgements}
This work was supported by the Australian Research Council through grant DP140102613,
and is  part of the program of study supported by the 
ARC Centre of Excellence for Mathematical \& Statistical Frontiers. At the beginning of the millennium, while attending international
workshops, 
I asked 3 eminent mathematicians about combining random matrix theory and the exceptional Jordan algebra for purposes of
being able to sample from (\ref{TX}). The first said it was not possible, as the theory was essentially abstract; the next told me it
would be  possible but he didn't know how; the third said it was possible and then spent the next period of time verbally explaining to me how to do it. Hence a special thanks to Eric Rains.


  \providecommand{\bysame}{\leavevmode\hbox to3em{\hrulefill}\thinspace}
\providecommand{\MR}{\relax\ifhmode\unskip\space\fi MR }
\providecommand{\MRhref}[2]{%
  \href{http://www.ams.org/mathscinet-getitem?mr=#1}{#2}
}
\providecommand{\href}[2]{#2}

 \end{document}